# Nucleosynthetic Pt isotope anomalies and the Hf-W chronology of core formation in inner and outer solar system planetesimals

Fridolin Spitzer[1*], Christoph Burkhardt[1], Francis Nimmo[2], Thorsten Kleine[1]

[1]Institut für Planetologie, University of Münster, Wilhelm-Klemm-Str. 10, 48149 Münster, Germany

[2]Department of Earth and Planetary Sciences, University of California Santa Cruz, Santa Cruz CA 95064, USA

*corresponding author: fridolin.spitzer@uni-muenster.de

## Abstract

The $^{182}$Hf-$^{182}$W chronology of iron meteorites provides crucial information on the timescales of accretion and differentiation of some of the oldest planetesimals of the Solar System. Determining accurate Hf-W model ages of iron meteorites requires correction for cosmic ray exposure (CRE) induced modifications of W isotope compositions, which can be achieved using *in-situ* neutron dosimeters such as Pt isotopes. Until now it has been assumed that all Pt isotope variations in meteorites reflect CRE, but here we show that some ungrouped iron meteorites display small nucleosynthetic Pt isotope anomalies. These provide the most appropriate starting composition for the correction of CRE-induced W isotope variations in iron meteorites from all major chemical groups, which leads to a ~1 Ma upward revision of previously reported Hf-W model ages. The revised ages indicate that core formation in non-carbonaceous (NC) iron meteorite parent bodies occurred at ~1–2 Ma after CAI formation, whereas most carbonaceous (CC) iron meteorite parent bodies underwent core formation ~2 Ma later. We show that the younger CC cores have lower Fe/Ni ratios than the earlier-formed NC cores, indicating that core formation under more oxidizing conditions occurred over a more protracted timescale. Thermal modeling of planetesimals heated by $^{26}$Al-decay reveals that this protracted core formation timescale is consistent with a higher fraction of water ice in CC compared to NC planetesimals, implying that in spite of distinct core formation timescales, NC and CC iron meteorite parent bodies accreted about contemporaneously within ~1 Ma after CAI formation, but at different radial locations in the disk.

# 1. Introduction

The fundamental isotopic dichotomy between *non-carbonaceous* (NC) and *carbonaceous* (CC) meteorites implies that planetesimal formation occurred in two spatially distinct areas of the solar accretion disk (Budde et al., 2016; Warren, 2011), which may have been separated through the rapid formation of proto-Jupiter (Kruijer et al., 2017) or a structured protoplanetary disk (Brasser and Mojzsis, 2020; Charnoz et al., 2021; Lichtenberg et al., 2021). Understanding the timescale of planetesimal formation in both reservoirs is, therefore, crucial for constraining the early dynamical evolution of the disk. Some of the most precise age constraints on planetesimal accretion may be obtained from application of the short-lived $^{182}$Hf-$^{182}$W system to "magmatic" iron meteorites (*e.g.*, Kleine et al., 2005). These are widely held to sample the metallic cores of differentiated planetesimals (Scott, 1972), and so the Hf-W ages of these irons provide the time of metal segregation in their parent bodies. These ages can, in turn, be linked to a parent body accretion age by thermal models of planetesimals heated by $^{26}$Al-decay (*e.g.,* Hilton et al., 2019; Kaminski et al., 2020; Kruijer et al., 2014b; Qin et al., 2008).

Determining Hf-W ages for iron meteorites requires correction for the effects of cosmic ray exposure (CRE), which induced secondary neutron capture reactions that can modify W isotope ratios (Masarik, 1997). Correction for these CRE-effects is commonly achieved using Pt or Os isotopes as the neutron dosimeter, and by assuming that all Pt and Os isotope variations reflect CRE (*e.g.*, Kruijer et al., 2013; Qin et al., 2015; Wittig et al., 2013; Worsham et al., 2017). Consistent with this, prior studies have found no evidence for nucleosynthetic Pt or Os isotope anomalies at the ±10 parts-per-million (ppm) level (*e.g.*, Hunt et al., 2017; Kruijer et al., 2017; Walker, 2012; Worsham et al., 2019). However, if present, even nucleosynthetic Pt or Os anomalies of <10 ppm would have a significant effect on the chronological interpretation of $^{182}$W data.

Prior studies using either Pt or Os isotopes to determine CRE-corrected $^{182}$W compositions have shown that CC iron meteorites have systematically younger Hf-W metal segregation ages than NC irons (Hilton et al., 2019; Kruijer et al., 2017, 2014b; Tornabene et al., 2020). This has been interpreted to reflect a slightly later accretion of CC compared to NC iron meteorite parent bodies (Hilton et al., 2019; Kruijer et al., 2017), which in turn has important implications for understanding the timescale over which the NC and CC reservoirs were separated. However, the difference in calculated Hf-W metal segregation ages could also reflect unresolved nucleosynthetic Pt isotope variability between NC and CC meteorites, which would then have led to systematically different pre-exposure (*i.e.*, unaffected by CRE) [182]W compositions by correction to a common Pt isotope composition. Thus, the reliable interpretation of Hf-W ages, and in particular of the Hf-W age difference between NC and CC irons, requires an assessment of whether there is any nucleosynthetic Pt isotope difference between the NC and CC reservoirs. However, until now, such an assessment has been hampered by the presence of super-imposed large CRE-effects on Pt isotopes, which makes identifying any much smaller nucleosynthetic isotope variations quite difficult.

Here we show that some ungrouped iron meteorites display only minor if any CRE-effects, but instead are characterized by small nucleosynthetic Pt isotope anomalies. This finding requires revision of pre-exposure [182]W compositions and Hf-W metal segregation ages for iron meteorites obtained in prior studies. The revised Hf-W ages are used to examine whether core formation in NC and CC irons occurred over distinct timescales, and to assess the chemical and physical processes affecting the timing of core formation in their parent bodies. Combined, these data are used to determine accretion ages of the NC and CC iron meteorite parent bodies and, hence, to assess planetesimal formation times in the inner and outer disk.

## 2. Samples and analytical methods

### 2.1. Sample selection

The ungrouped iron meteorites ILD 83500 and Babb's Mill (Troost's Iron) (BM.18490) belong to the South Byron Trio (SBT) and were selected for this study because their small recovered masses of 2.5 kg and 2.7 kg suggest that their pre-atmospheric sizes may have been too small for significant CRE-induced neutron capture reactions and because a prior study did not find any resolvable CRE-induced Os isotope variations in these samples (Hilton et al., 2019). The two other ungrouped iron meteorites are Guffey (USNM 4832) and Hammond (USNM 471), which as part of a larger systematic study on ungrouped irons were found to have identical Pt isotope compositions as ILD 83500 and Babb's Mill.

In addition to the iron meteorites, two CB chondrites (Gujba, Isheyevo) and metal separates from three equilibrated ordinary chondrites (Butsura, NWA 6630, NWA 6629) were selected for this study. The choice of these particular samples mainly reflects that they are metal-rich, rendering it possible to analyze pure metal samples. Moreover, the comparison of CB and ordinary chondrites may help assess whether there is any nucleosynthetic Pt isotope difference between NC and CC meteorites. In general, CRE ages of chondrites are short compared to those of most iron meteorites (*e.g.*, Herzog and Caffee, 2013), making chondrites promising targets for identifying nucleosynthetic Pt isotope anomalies in meteorites.

## 2.2. Analytical techniques

Pieces of the iron meteorites and the CB chondrite Isheyevo were cut using a diamond saw and polished with abrasives (SiC). The CB chondrite Gujba samples were individual metal chondrules separated from a disaggregated sample. Metal separates from the two ordinary chondrites NWA 6630 and NWA 6629 were left-over material from a Hf-W study (Hellmann et al., 2019), and the metal separate from Butsura was newly prepared by metal-silicate separation using a hand magnet (see Hellmann et al. 2019 for details). All samples were cleaned by ultrasonication in ethanol or acetone to remove any remaining adhering dust, so that clean metal samples were obtained.

Individual samples weighing between 0.05−0.6 g were digested in 6M HCl (+trace $HNO_3$) in Teflon beakers at 130 °C on a hot plate for at least 24 h. Upon complete dissolution of the iron meteorites, an aliquot corresponding to ~50 mg sample was taken for Pt isotope analyses, while the remainder was processed for Mo and W isotope analyses. For the chondrites, the resulting solutions were centrifuged to remove any undissolved silicates (which contain virtually no Pt), and the supernatant was used for the Pt isotope measurements. For these samples, no Mo and W isotope measurements were made.

The chemical separation and mass spectrometry procedures for Pt, Mo, and W used in this study followed our established methods and are described in detail in the Supplementary Material. The isotope data are corrected for mass bias by internal normalization using the exponential law, and are reported in the $\varepsilon$-notation as parts-per-10,000 deviations from terrestrial standard values. The accuracy and reproducibility of the isotope measurements were assessed by repeated analyses of separate digestions of the NIST 129c steel (Tables S1-S3).

## 3. Results

The four ungrouped irons show well-resolved $\varepsilon^i$Mo excesses relative to the terrestrial standard (Table 1), and plot on the CC-line in the $\varepsilon^{94}$Mo-$\varepsilon^{95}$Mo diagram (Fig. 1), indicating that they belong to the CC group of meteorites (*e.g.*, Kruijer et al., 2017). The Mo isotopic compositions of Babb's Mill and ILD 83500 determined in this study are in excellent agreement with previously reported results for the same meteorites obtained by thermal ionization mass spectrometry (Hilton et al., 2019).

The W isotope data for the four ungrouped irons are provided in Table S4. As for Mo, the W isotopic data for Babb's Mill and ILD 83500 are in excellent agreement with those reported by Hilton et al. (2019). These two samples exhibit slightly more negative $\varepsilon^{182}$W values (-3.28) than the other two irons of this study, Guffey (-3.10) and Hammond (-3.13). All four irons display small positive $\varepsilon^{183}$W anomalies of ~0.15, consistent with their CC heritage (*e.g.*, Kruijer et al., 2017).

The four ungrouped iron meteorites have indistinguishable Pt isotope compositions and are characterized by negative $\varepsilon^{192}$Pt, $\varepsilon^{194}$Pt, and $\varepsilon^{196}$Pt (Table 2, Table S5). This contrasts with the positive $\varepsilon^{192}$Pt, $\varepsilon^{194}$Pt, and $\varepsilon^{196}$Pt commonly observed for iron meteorites having CRE-induced Pt isotope shifts, implying that the negative anomalies of the ungrouped irons of this study are not due to CRE.

The two CB chondrites have indistinguishable Pt isotope compositions, which partly overlap with the composition of the four ungrouped irons of this study (Fig. 2). The measured $\varepsilon^{192}$Pt value of Isheyevo is more positive than those of the irons, but owing to an unusually large Os interference correction the $\varepsilon^{192}$Pt of this sample is also more uncertain and within its larger uncertainty also overlaps with the composition of the ungrouped irons. Thus, the Pt isotope composition of the CB chondrites is similar to those of the four ungrouped irons, but due to the larger uncertainty on the chondrite data their anomalies are less well resolved from the standard. Finally, none of the three ordinary chondrites displays clearly resolved Pt isotope anomalies, but compared to the ungrouped irons of this study, exhibit slightly more positive $\varepsilon^{192}$Pt, $\varepsilon^{194}$Pt, and $\varepsilon^{196}$Pt. As will be shown below, these differences are consistent with small CRE-induced Pt isotope anomalies in the ordinary chondrites.

## 4. Nucleosynthetic Pt isotope anomalies in meteorites

### 4.1. Identification of nucleosynthetic Pt isotope anomalies in ungrouped irons

Platinum isotope variations in meteorites predominantly reflect CRE-induced $n$-capture reactions (Kruijer et al., 2013; Leya and Masarik, 2013; Wittig et al., 2013), where the dominant reactions are: $^{195}$Pt$(n,\gamma)^{196}$Pt, $^{191}$Ir$(n,\gamma)^{192}$Ir$(\beta^-)^{192}$Pt, and $^{193}$Ir$(n,\gamma)^{194}$Ir$(\beta^-)^{194}$Pt (Leya and Masarik, 2013). Evidently, CRE-induced anomalies on $^{192}$Pt and $^{194}$Pt depend on the sample's Ir/Pt ratio, whereas the anomalies on $^{196}$Pt do not. Importantly, CRE-induced reactions can only result in elevated $\varepsilon^{192}$Pt and $\varepsilon^{194}$Pt, because both $^{192}$Pt and $^{194}$Pt are produced by $n$-capture on Ir and subsequent $\beta^-$-decay. Also, CRE can only lead to positive $\varepsilon^{196}$Pt, because the $^{196}$Pt/$^{195}$Pt ratio increases through $n$-capture on $^{195}$Pt to produce $^{196}$Pt. Thus, the negative $\varepsilon^{192}$Pt, $\varepsilon^{194}$Pt, and $\varepsilon^{196}$Pt observed for some of the samples of this study cannot reflect CRE relative to the Pt isotope composition of the standard.

To assess the origin of the negative $\varepsilon^{192}$Pt, $\varepsilon^{194}$Pt, and $\varepsilon^{196}$Pt values, it is useful to not only normalize the Pt isotope data to $^{198}$Pt/$^{195}$Pt (termed '8/5'), but also to other Pt isotope ratios, such as $^{196}$Pt/$^{195}$Pt ('6/5') and in particular $^{198}$Pt/$^{194}$Pt ('8/4'). The latter has so far not been used in prior studies, mainly because in most samples the abundance of $^{194}$Pt is modified by CRE. However, for samples unaffected by CRE, the $^{198}$Pt/$^{194}$Pt-normalization is useful to assess whether the observed Pt isotope variations may reflect mass-independent isotope fractionation, for instance by the nuclear field shift (NFS). In fact, prior studies have argued that the NFS effect may be responsible for some of the isotope anomalies observed among meteorites. However, although a NFS fractionation may result in negative $\varepsilon^{192}$Pt (8/4), it would then lead to positive $\varepsilon^{195}$Pt (8/4) and $\varepsilon^{196}$Pt (8/4), which is inconsistent with the observed Pt isotope pattern of the four ungrouped irons (Fig. 3). This observation holds for the other normalizations of the Pt isotope data (Fig. 3), indicating that the Pt isotope anomalies of the ungrouped irons of this study cannot be due to the NFS effect. Instead, regardless of which normalization is used, the Pt isotope pattern of these samples is always consistent with a nucleosynthetic origin and, more specifically, with an $s$-process deficit or $r$-process excess relative to the terrestrial standard (Fig. 3). We, therefore, interpret the negative $\varepsilon^{192}$Pt, $\varepsilon^{194}$Pt, and $\varepsilon^{196}$Pt (8/5-normalization) of the four ungrouped irons of this study to be nucleosynthetic in origin. This is the first identification of nucleosynthetic Pt isotope anomalies in meteorites.

It is noteworthy that prior work on Pt isotopes in iron meteorites has not found resolvable nucleosynthetic Pt isotope anomalies, although analytical techniques with similar precision to this study were used (*e.g.*, Hunt et al., 2017; Kruijer et al., 2013). This is because the Pt isotope anomalies in iron meteorites are predominantly governed by CRE-induced shifts, which typically are much larger than the nucleosynthetic Pt isotope anomalies reported in this study. For instance, many iron meteorites display $\varepsilon^{196}$Pt excesses of between ~0.2 and ~0.4, and corresponding $\varepsilon^{192}$Pt excesses of up to several tens of $\varepsilon$ (Kruijer et al., 2014b, 2013). For such samples, the much smaller nucleosynthetic Pt isotope anomalies are completely overprinted by CRE-induced shifts. The identification of nucleosynthetic Pt isotope anomalies in this study, therefore, results from the discovery of rare iron meteorites with very small or absent CRE-induced Pt isotope shifts. Of note, the true pre-exposure $\varepsilon^{192}$Pt, $\varepsilon^{194}$Pt, and $\varepsilon^{196}$Pt values of the irons may be more negative than the measured values, which may still be partly affected by CRE. We will return to this issue in more detail below.

Given the presence of nucleosynthetic Pt isotope anomalies in iron meteorites, it might be expected that the neighboring element Os also shows nucleosynthetic anomalies. However, so far, no nucleosynthetic Os isotope anomalies in iron meteorites have been identified (Walker, 2012; Wittig et al., 2013; Worsham et al., 2017). As for Pt, non-radiogenic Os isotope variations among iron meteorites predominantly reflect CRE-induced isotope shifts, making the identification of nucleosynthetic Os isotope anomalies quite difficult. Hilton et al. (2019) showed that two of the irons of this study (Babb's Mill, ILD 83500) have no resolvable Os isotope anomalies, indicating that these samples are only minimally affected by CRE. However, within the analytical uncertainty of the Os isotope measurements, these samples may still have small nucleosynthetic Os anomalies of similar magnitude as the Pt isotope anomalies identified in the present study. More specifically, assuming a maximum expansion of the errors, Babb's Mill and ILD 83500 may have slightly positive $\mu^{189}$Os values of up to 15, and slightly negative $\mu^{190}$Os values of down to –11 (Hilton et al., 2019), which would be consistent with the expected pattern of nucleosynthetic Os isotope anomalies. Thus, although these samples display no resolved Os isotope anomalies, their Os isotope compositions are not inconsistent with the identification of small nucleosynthetic Pt isotope anomalies in this study.

*4.2. Distinct Pt isotope compositions for non-carbonaceous and carbonaceous meteorites?*

Most elements that exhibit nucleosynthetic isotope anomalies among bulk meteorites also display systematic differences between the NC and CC reservoirs. This includes Ca, Ti, Cr, Ni, Mo, Ru, and W, all of which have systematically different isotopic compositions between NC and CC meteorites [see Kleine et al. (2020) for a review]. In general, compared to the terrestrial composition, the isotope anomalies in CC meteorites are usually larger than those in NC meteorites. Moreover, for elements beyond the Fe-peak, Earth defines an endmember isotopic composition compared to all meteorites. As such, it might be expected that the nucleosynthetic Pt isotope anomalies also increase in the order Earth < NC < CC towards a greater *r*-process excess or *s*-process deficit (*i.e.*, more negative $\varepsilon^{192}$Pt, $\varepsilon^{194}$Pt, and $\varepsilon^{196}$Pt).

To assess whether there is any nucleosynthetic Pt isotope difference between NC and CC meteorites, we compiled literature Pt isotope data (Table S6) for iron meteorites for which CRE-induced isotope shifts are minor to absent ($\varepsilon^{196}$Pt < 0.10) (Fig. 4). Many of these samples have slightly negative $\varepsilon^{196}$Pt and some also have slightly negative $\varepsilon^{194}$Pt, both of which are, however, not resolved from zero. Moreover, neither of these samples display resolved negative $\varepsilon^{192}$Pt values, unlike the four ungrouped irons of this study. Among the NC irons, Guadalupe y Calvo (IIAB iron) has the lowest $\varepsilon^{196}$Pt and $\varepsilon^{194}$Pt, which agree with the composition of the four ungrouped CC irons of this study. However, Guadalupe y Calvo has a more positive $\varepsilon^{192}$Pt than these samples, but again this difference is not resolved. Altogether, although NC irons tend to have slightly more positive Pt isotope anomalies than CC irons, there is no resolved difference between these two groups of irons and the Pt isotope variations among the irons with $\varepsilon^{196}$Pt < 0 are equally consistent with small nucleosynthetic variations or small CRE-induced effects.

Another way to assess potential nucleosynthetic Pt isotope differences between NC and CC meteorites is to examine ordinary chondrites. This is because compared to iron meteorites, chondrites typically have much shorter CRE times (*e.g.*, Herzog and Caffee, 2013), and so CRE-induced isotope shifts in chondrites are smaller. The three ordinary chondrites of this study have more positive $\varepsilon^{192}$Pt, $\varepsilon^{194}$Pt, and $\varepsilon^{196}$Pt than the four ungrouped CC irons, and also than some of the NC irons (Fig. 4). It is unlikely that the latter difference reflects nucleosynthetic Pt isotope variations, because ordinary chondrites and IVA iron meteorites have indistinguishable Mo and Ru isotope anomalies (Kleine et al. (2020) and references therein), yet the IVA iron with the smallest CRE-induced Pt isotope effects (Gibeon) has more negative $\varepsilon^{196}$Pt and $\varepsilon^{194}$Pt than the ordinary chondrites of this study. The slightly elevated $\varepsilon^{192}$Pt, $\varepsilon^{194}$Pt, and $\varepsilon^{196}$Pt values of the ordinary chondrites, therefore, most likely reflect CRE. This may seem unexpected, given the short CRE times of ordinary chondrites, but is consistent with the higher expected fluence of thermal neutron in chondrites compared to iron meteorites (Leya and Masarik, 2013). As a result, the ordinary chondrite data of this study also provide no clear evidence for distinct Pt isotope compositions of NC and CC meteorites.

A final possibility to address this issue is by back-projection of the measured Pt isotope compositions of irradiated iron meteorites to a common intersection point. In plots of $\varepsilon^{192}$Pt versus $\varepsilon^{196}$Pt and $\varepsilon^{194}$Pt versus $\varepsilon^{196}$Pt (Fig. S1-S4), CRE-induced shifts follow straight lines whose slopes depend on a sample's Ir/Pt ratio. Thus, examining whether or not NC and CC irons have a common intersection point in the $\varepsilon^{192}$Pt-$\varepsilon^{196}$Pt and $\varepsilon^{194}$Pt-$\varepsilon^{196}$Pt diagrams may help identifying any potential nucleosynthetic Pt isotope difference between the NC and CC reservoirs. Figure 4 shows that the Pt isotope data of all irons, including NC and CC samples, are consistent with a common starting point at around the composition measured for the four ungrouped CC irons of this study (Fig. 4b). This observation shows once again that with the current data set there is no resolvable nucleosynthetic Pt isotope difference between NC and CC meteorites, and it also suggests that, if present, any such difference must be small, because NC and CC meteorites have a common focal point in the $\varepsilon^{192}$Pt-$\varepsilon^{196}$Pt and $\varepsilon^{194}$Pt-$\varepsilon^{196}$Pt diagrams. A corollary of this observation is that the Pt isotope composition of the four ungrouped irons of this study provides a good estimate of the pre-exposure Pt isotope composition of iron meteorites in general, because otherwise the back-projection of other irradiated samples would intersect at a composition characterized by more negative $\varepsilon^{192}$Pt, $\varepsilon^{194}$Pt, and $\varepsilon^{196}$Pt. We, therefore, interpret the mean Pt isotope composition of the ungrouped irons of this study to provide the current best estimate for the pre-exposure Pt isotope composition of iron meteorites in general, including irons from the NC and CC reservoirs.

## 5. Implications for Hf-W chronology of iron meteorites

### 5.1. Revised pre-exposure $\varepsilon^{182}$W of iron meteorites

The identification of nucleosynthetic Pt isotope anomalies has important implications for the Hf-W chronology of core formation in iron meteorite parent bodies. This is because calculating accurate Hf-W model ages requires knowledge of pre-exposure $\varepsilon^{182}$W values of iron meteorites, which may either be obtained from the *y*-axis intercept of empirical $\varepsilon^{182}$W-$\varepsilon^{196}$Pt correlation lines for each group of irons (Kruijer et al., 2013), or by correcting individual samples using the following equation (Kruijer et al., 2017):

$$\varepsilon^{182}W_{\text{pre-exposure}} = \varepsilon^{182}W_{\text{measured}} - \varepsilon^{196}Pt_{\text{measured}} \times (-1.320 \pm 0.055) \tag{1}$$

Both approaches assume that the pre-exposure $\varepsilon^{196}$Pt of iron meteorites is zero (*i.e.*, identical to the composition of the terrestrial standard). However, as shown here, a more appropriate pre-exposure Pt isotope composition for both NC and CC irons is provided by the four ungrouped CC irons of this study, which have a mean $\varepsilon^{196}Pt = -0.06 \pm 0.01$ (Table 2). Using this value instead of zero results in an upward revision of pre-exposure $\varepsilon^{182}W$ values for both NC and CC irons by ~0.08 (*i.e.*, 0.06 × 1.32, eq. 1). Accordingly, individual samples should be corrected using the following, updated equation:

$$\varepsilon^{182}W_{\text{pre-exposure}} = \varepsilon^{182}W_{\text{measured}} - [\varepsilon^{196}Pt_{\text{measured}} + 0.06(\pm 0.01)] \times (-1.320 \pm 0.055) \qquad (2)$$

The updated pre-exposure values for the major magmatic iron meteorite groups are shown in Figure 5. Note that the $\varepsilon^{182}W$ of the ungrouped irons from this study were not corrected, because their Pt isotope compositions show no evidence for significant CRE-effects.

One important implication of the results of this study is that the ~0.2 $\varepsilon^{182}W$ difference and, hence, ~2 Ma model age difference between NC and CC irons (Kruijer et al., 2017) cannot reflect unresolved nucleosynthetic Pt isotope variability. This is because for a pre-exposure $\varepsilon^{196}Pt = -0.06$ of the CC irons, the pre-exposure $\varepsilon^{196}Pt$ of NC irons would have to be *ca.* -0.20 to account for the $\varepsilon^{182}W$ difference between these two groups of irons. However, as noted above, if anything, NC meteorites would have smaller nucleosynthetic Pt isotope anomalies than CC irons. Moreover, for all elements investigated thus far, CC meteorites have larger nucleosynthetic isotope anomalies than NC meteorites (*e.g.*, Kleine et al., 2020), and there is no reason to assume that this order would be reversed for Pt isotopes. Thus, the results of this study confirm that CC irons have systematically more radiogenic $\varepsilon^{182}W$ values than NC irons.

Another important result of this study is that the Hf-W age differences among the NC iron meteorite groups also do not reflect nucleosynthetic Pt isotope variability. If such variability were to exist, the CRE-correction of measured $\varepsilon^{182}W$ values to a common $\varepsilon^{196}Pt$ may have led to apparent variations in pre-exposure $\varepsilon^{182}W$ values among the NC iron meteorite groups. However, the two NC iron groups with the largest $\varepsilon^{182}W$ difference, the IC and IVA irons, have very similar nucleosynthetic Mo isotope anomalies (e.g., $\varepsilon^{94}Mo = 0.90$ or 0.79, respectively) (Spitzer et al., 2020). Given that the nucleosynthetic isotope variability among meteorites is much larger for Mo than for Pt (*i.e.*, by a factor of ~20), any significant nucleosynthetic Pt isotope difference among the NC irons can therefore be excluded. The variations in pre-exposure $\varepsilon^{182}$W values among NC iron meteorites are, therefore, radiogenic in origin.

## 5.2. Origin of distinct core formation times in NC and CC irons

A model age of metal segregation can be calculated as the time of Hf–W fractionation from an unfractionated reservoir with chondritic Hf/W using the following equation (*e.g.*, Kleine and Walker, 2017):

$$\Delta t_{CAI} = -\frac{1}{\lambda} \times ln \left[ \frac{\varepsilon^{182}W_{sample} - \varepsilon^{182}W_{chondrites}}{\varepsilon^{182}W_{SSI} - \varepsilon^{182}W_{chondrites}} \right] \tag{3}$$

where $\varepsilon^{182}$W$_{sample}$ is the W isotope composition of an iron meteorite, $\varepsilon^{182}$W$_{chondrites}$ is the composition of carbonaceous chondrites (-1.9 ± 0.1) (Kleine et al., 2009), ($\varepsilon^{182}$W)$_{SSI}$ is the Solar System initial (-3.49 ± 0.07) obtained from CAIs (Kruijer et al., 2014a), and $\lambda$ is the decay constant of $^{182}$Hf of 0.0778 ± 0.0015 Ma$^{-1}$ (Vockenhuber et al., 2004). Using these parameters and the revised pre-exposure $\varepsilon^{182}$W values from this study results in revised model ages that are ~0.8 Ma younger than previously calculated ages and which are ~1–2 Ma for NC irons and ~3–4 Ma after CAI formation for most CC irons, respectively (Table 3, Fig. 5). Only samples of the SBT (including ILD 83500 and Babb's Mill) display somewhat lower pre-exposure $\varepsilon^{182}$W values than the other CC irons, indicating that core formation in their parent body occurred ~1 Ma earlier and at about the same time as in some NC iron meteorite parent bodies (Fig. 5).

Calculating model ages for NC and CC iron meteorite parent bodies using eq. 3 relies on the assumption that prior to core formation these bodies had the same Hf/W ratios and, therefore, followed the same chondritic Hf-W isotopic evolution as determined based on Hf-W data for carbonaceous chondrites (Budde et al., 2018; Kleine et al., 2004, 2002; Kruijer et al., 2014a). However, Hellmann et al. (2019) showed that the precursor material of ordinary chondrites, which belong to the NC meteorites, was characterized by a lower Hf/W ratio ($^{180}$Hf/$^{184}$W ≈ 0.7) than carbonaceous chondrites ($^{180}$Hf/$^{184}$W ≈ 1.35). A $^{180}$Hf/$^{184}$W ratio of 0.7 would lead to a present-day $\varepsilon^{182}$W of -2.67, and results in Hf-W model ages for NC irons that are ~1 Ma younger (eq. 3) than the ages calculated assuming a carbonaceous chondrite-like Hf/W (Hellmann et al., 2019). Nevertheless, the model age for the IC and IIAB irons, which are the NC irons with the least radiogenic W isotopic compositions, are still ~1 Ma older than the ages for the CC irons. Thus, different Hf/W ratios can account for some but not the entire difference in model ages between NC and CC irons, suggesting that core formation in CC irons occurred at least ~1 Ma later than in NC irons.

The later core formation in CC iron meteorite parent bodies may either reflect later accretion or a delayed onset of melting and core formation compared to NC iron meteorite parent bodies. For instance, CC bodies likely accreted at or beyond the snowline (*e.g.*, Kruijer et al., 2017), and so CC iron meteorite parent bodies may have incorporated water ice, which may have delayed the onset of melting predominantly by lowering the concentration of heat-producing $^{26}$Al (see below). The presence of water ice would have also led to oxidation, and so core formation in CC bodies should have occurred under more oxidizing conditions than in NC bodies. This in turn should have resulted in smaller metallic cores in CC compared to NC bodies, because more Fe remained in the mantle of CC bodies. The core sizes of iron meteorite parent bodies can be estimated from the bulk HSE abundances inferred for each core, because the HSE partition nearly quantitatively into the core (*e.g.*, Tornabene et al., 2020). However, calculating core sizes in this manner requires assumptions about the bulk composition of iron meteorite parent bodies prior to differentiation, which were likely different for NC and CC bodies and may have also varied among different members of either the NC or CC group. In spite of this uncertainty, the core sizes inferred from bulk HSE abundances tend to be smaller for CC compared to NC objects, although there is overlap in the estimated core sizes (*e.g.*, Rubin, 2018).

Another way to examine the conditions of core formation that is independent of assumed bulk compositions is to use variations in the Fe/Ni ratios of the bulk cores. At the low-pressure conditions of core formation in iron meteorite parent bodies, Ni is strongly siderophile, and so almost all Ni partitions into the core (*e.g.*, Fischer et al., 2015). Thus, the Fe/Ni ratio of the core is a measure of how much Fe remained in the mantle, such that a lower Fe/Ni ratio of the core indicates core formation under more oxidizing conditions. The only other elements besides Fe that occur in significant amounts in the metal cores of iron meteorite parent bodies are Ni, S, and P, whereas all other elements typically have concentrations below 1 wt.% (*e.g.*, Chabot, 2004; McCoy et al., 2011; Walker et al., 2008). Thus, the bulk core Fe content can be estimated provided that the bulk Ni, S, and P contents of the core are known. To this end we used the bulk S and P contents of iron meteorite cores inferred in prior studies, and calculated the bulk Ni content of each core using the Ni concentrations of the least evolved member of each group together with the solid metal-liquid melt partition coefficient of Ni at the appropriate S content (Table S7). The bulk Fe content is then obtained by mass balance. The results show that CC irons have systematically lower Fe/Ni ratios than NC irons, indicating that core formation in CC bodies occurred under more oxidizing conditions than in NC bodies (Fig. 6a). This is, at least qualitatively, consistent with a larger water ice fraction in CC compared to NC bodies. As such, the more radiogenic $\varepsilon^{182}$W and later inferred core formation time of CC compared to NC iron meteorite parent bodies may at least partly reflect chemical differences that led to a delayed onset of core formation.

To assess these observations more quantitively, we calculated the thermal evolution of plane-tesimals heated internally by $^{26}$Al-decay as a function of water ice fraction in the bulk objects. The model is described in detail in the supplement. In brief, it uses a similar approach to the conduction model of Kruijer et al. (2017) except that the thermal properties and heat production rate are modified based on the ice mass fraction present. It also accounts in an approximate way for enhanced heat transfer via porous water convection using an approach applied to hydrothermal systems (see supplement for details). The results of the thermal model show that the presence of water ice significantly delays the onset of melting and, hence, the time of core formation. For instance, for an accretion time of 0.5 Ma, bodies containing 20-30 wt.% water ice would undergo core formation ~0.5-1 Ma later than bodies containing no water ice (Fig. 7a). This time difference increases if the effects of water convection are also considered, in which case core formation is delayed by ~1-2 Ma (Fig. 7b). Thus, in addition to different precursor Hf-W ratios and accretion timescales, the distinct core formation times of NC and CC iron meteorite parent bodies may also be accounted for by different water ice fractions in these bodies.

### 5.3. Accretion ages of NC and CC iron meteorite parent bodies

Several studies have shown that thermal models like those of this study can be used to link the Hf-W model ages of core formation to parent body accretion times (*e.g.*, Hilton et al., 2019; Kruijer et al., 2014b; Qin et al., 2008). Unlike in these previous works, the thermal model of this study also considers the effect of water ice on the time of differentiation and thus allows a more realistic assessment of the accretion times of the CC iron meteorite parent bodies. This in turn is essential for evaluating whether or not the parent bodies of NC and CC irons formed at the same time.

One major problem for determining the accretion time of CC iron meteorite parent bodies is that their initial water ice fractions are unknown. As is evident from their subchondritic Ga/Ni and Ge/Ni ratios, these bodies are all variably depleted in volatile elements (Scott and Wasson, 1975). These depletions extend to the S contents inferred for each CC core, which are also variable and are roughly correlated with the degree of volatile depletion (Hilton et al., 2020, 2019; Tornabene et al., 2020; Walker et al., 2008). The S content of the bulk core exerts a strong control on the melting temperature of the metal and, hence, the core formation time (Kruijer et al., 2014b). Consequently, if the variable volatile depletions of the CC irons were to reflect their initial compositions prior to parent body differentiation, then these bodies should have undergone core formation at different times. However, despite their variable S contents, all CC irons (except perhaps the SBT) have indistinguishable $\varepsilon^{182}$W values (Fig. 6b), indicating that core formation occurred at about the same time. This suggests that the volatile element depletions were established after core formation and that, therefore, the CC iron meteorite parent bodies were volatile-richer than they are today. Thus, the volatile element depletion of CC irons does not imply that these bodies did not contain water ice initially.

The CC irons are isotopically linked to carbonaceous chondrites, some or all of which incorporated water ice. The water contents of carbonaceous chondrites vary from concentrations of ~8 wt.% for CV chondrites up to ~20-30 wt.% for CI chondrites (Alexander, 2019a). However, due to terrestrial contamination, the indigenous water contents of carbonaceous chondrites may have been ~2 times lower (Vacher et al., 2020). The lower water concentrations of most carbonaceous chondrites compared to CI chondrites may reflect water loss during parent body metamorphism (*e.g.*, Vacher et al., 2020) or lower indigenous contents due to the incorporation of a smaller fraction of water-rich, CI-like matrix (Alexander, 2019a; Hellmann et al., 2020). It is noteworthy that even the CI chondrites contain less water ice than inferred from the theoretical water-rock ratio of the solar nebula of 1.2, which corresponds to ~55 wt.% water ice (Krot et al., 2015). This suggests either that water ice accretion was not efficient (Cyr et al., 1998) or that the water-rock ratio was lower than the theoretical solar value because some O was bound to C (Alexander, 2019a). Together, the water contents of carbonaceous chondrites do not provide unambiguous estimates for the initial water ice fractions in CC iron meteorite parent bodies, but they nevertheless show that water contents of up to ~20-30 wt.% are not implausible. Moreover, carbonaceous chondrites represent a relatively late generation of planetesimals, whose accretion might have been triggered by dust enrichment in a pressure bump outside of Jupiter's orbit (Desch et al., 2018). By contrast, the CC iron meteorite parent bodies formed earlier and possibly by a different mechanism, which may have involved dust enrichment at or beyond the snowline (Drążkowska and Alibert, 2017). As such, the parent bodies of CC irons may have initially contained higher water ice fractions than the CI chondrites. A final important observation is that, as noted above, the lower Fe/Ni ratios of the CC compared to NC cores (Fig. 6a) indicate core formation under more oxidizing conditions, suggesting that the parent bodies of the CC irons accreted a larger fraction of water ice than those of the NC irons.

The CC irons (except the SBT) have indistinguishable Hf-W model ages averaging at 3.3±0.6 (2 s.d.) Ma after CAI formation. The results of the thermal model show that for this core formation time and assuming a water ice fraction of zero, accretion would have occurred at ~1.2 Ma after CAI formation, consistent with results of prior studies using a similar thermal model (Hilton et al., 2019; Kruijer et al., 2017). However, since the CC iron meteorite parent bodies likely incorporated some water ice, accretion must have occurred earlier. For instance, assuming a water ice fraction of ~20-30 wt.% results in an accretion age of ~0.7 Ma, while assuming the theoretical water-rock ratio of the solar nebula returns an accretion age of <0.5 Ma (Fig. 7a). Within this context, the slightly earlier core formation time of the SBT compared to the other CC irons may reflect either a lower water ice fraction or an earlier accretion time. These examples highlight that the CC iron meteorite parent bodies likely accreted well within ~1 Ma and, depending on their initial water ice content, perhaps even within the first ~0.5 Ma after CAI formation.

The higher Fe/Ni ratios inferred for the NC iron meteorite cores indicate more reducing conditions of core formation and, hence, lower initial water ice fractions in these bodies. This is also consistent with the low water contents of ordinary and enstatite chondrites (Alexander, 2019b; Piani et al., 2020). Thus, the accretion ages of NC iron meteorite parent bodies can be inferred from the thermal evolution of bodies containing no water ice. As noted by Kruijer et al. (2017), the accretion ages are most reliably determined using Hf-W model ages of the volatile undepleted NC irons (*i.e.*, group IC and IIAB) because owing to their high initial S concentrations, core formation occurred as a single event of metal segregation. By contrast, for the volatile-depleted NC irons (*e.g.*, group IVA), core formation may have involved two stages of metal segregation, such that the accretion time cannot easily be linked to a specific core formation age. In fact, Kruijer et al. (2014b) argued that in spite of their different Hf-W model ages, all NC iron meteorite parent bodies accreted at about the same time. The Hf-W model ages of the IC and IIAB irons using the revised pre-exposure $\varepsilon^{182}W$ values of this study, and assuming a bulk carbonaceous chondrite-like Hf/W, are 1.0±0.7 Ma and 1.4±0.6 Ma, respectively. These ages correspond to an accretion time of ~0.5 Ma after CAI formation in the ice-free case (Fig. 7). If the lower Hf/W ratio of the precursor material of ordinary chondrites is used in the calculation (see above), the Hf-W model ages change to 2.0±1.3 Ma and 3.0±1.3 Ma, respectively. For these younger core formation ages, the inferred accretion times change to ~1 Ma after CAI formation (Fig. 7). Thus, as for the CC irons, the Hf-W data show that accretion of NC iron meteorite parent bodies occurred within ~1 Ma after CAI formation and possibly earlier.

In summary, the distinct core formation times of NC and CC iron meteorite parent bodies can be fully accounted for by a higher water ice fraction in the latter, which delayed the onset of melting and core formation. The Hf-W data show that the parent bodies of both NC and CC irons accreted within ~1 Ma after CAI formation, and currently do not allow resolving differences in the accretion times of these objects. As such, these data reveal that rapid planetesimal formation occurred about concurrently in the NC and CC reservoirs and involved the formation of chemically distinct objects, which underwent core formation over distinct timescales. This corroborates earlier conclusions that NC and CC planetesimals formed in spatially separated regions and, hence, distinct radial locations in the solar accretion disk.

## 6. Conclusions

The identification of small nucleosynthetic Pt isotope anomalies in some ungrouped iron meteorites has important implications for the Hf-W chronology of iron meteorites, because Pt isotopes are widely used as a neutron dosimeter for quantifying CRE-induced modifications of W isotope compositions. The correction for nucleosynthetic Pt isotope anomalies leads to Hf-W model ages for iron meteorites that are ~1 Ma younger than previously calculated. The revised Hf-W model ages indicate that core formation in NC iron meteorite parent bodies occurred at ~1–2 Ma after CAI formation, whereas core formation in most CC iron meteorite parent bodies occurred ~2 Ma later. We suggest that these different core formation timescales reflect a higher water ice fraction in the CC iron meteorite parent bodies, which led to a delayed onset of melting and metal segregation by lowering the concentration of heat-producing $^{26}$Al and by enhancing heat transfer via porous water convection. Results of a thermal model of bodies heated internally by $^{26}$Al-decay show that the NC and CC iron parent bodies accreted within ~1 Ma after CAI formation, implying that rapid planetesimal formation occurred about concurrently in the NC and CC reservoirs and resulted in the formation of chemically distinct objects. The modeling results also demonstrate that uncertainties in the Hf/W ratio of the precursor bodies of NC iron meteorites and in the initial water ice fraction of CC iron meteorite parent bodies currently preclude determination of more precise accretion timescales and whether or not one group of bodies formed earlier than the other.

**Acknowledgements**

We are grateful to Natasha Almeida (National History Museum, London) and Tim McCoy (Smithsonian Institution, Washington D.C.) for providing samples for this study. Comments by two anonymous reviewers and the efficient editorial handling by Fred Moynier are gratefully acknowledged. This study was funded by the Deutsche Forschungsgemeinschaft (DFG, German Research Foundation) – Project-ID 263649064 – TRR 170. This is TRR 170 pub. no. 141.

Table 1

Molybdenum isotopic data for ungrouped iron meteorites.

| Sample | ID (Mo-IC) | N | $\varepsilon^{92}Mo_{meas.}$ (± 95% CI) | $\varepsilon^{94}Mo_{meas.}$ (± 95% CI) | $\varepsilon^{95}Mo_{meas.}$ (± 95% CI) | $\varepsilon^{97}Mo_{meas.}$ (± 95% CI) | $\varepsilon^{100}Mo_{meas.}$ (± 95% CI) | Reference |
|---|---|---|---|---|---|---|---|---|
| Babb's Mill (Troost's Iron) | UI-01 | 7 | 1.89 ± 0.14 | 1.31 ± 0.10 | 1.07 ± 0.08 | 0.51 ± 0.07 | 0.55 ± 0.10 | this study |
| Babb's Mill (Troost's Iron) | | 5 | - | 1.32 ± 0.09 | 1.04 ± 0.05 | 0.46 ± 0.03 | - | Hilton et al. (2019) |
| Guffey | UI-20 | 6 | 1.82 ± 0.17 | 1.42 ± 0.11 | 1.03 ± 0.05 | 0.55 ± 0.05 | 0.60 ± 0.10 | this study |
| Hammond | UI-21 | 6 | 2.28 ± 0.16 | 1.76 ± 0.09 | 1.29 ± 0.03 | 0.65 ± 0.07 | 0.66 ± 0.08 | this study |
| ILD 83500 | UI-22 | 6 | 1.71 ± 0.16 | 1.20 ± 0.14 | 1.01 ± 0.08 | 0.53 ± 0.05 | 0.64 ± 0.10 | this study |
| ILD 83500 | | 4 | - | 1.21 ± 0.13 | 0.99 ± 0.09 | 0.50 ± 0.02 | - | Hilton et al. (2019) |

N is the number of measurements per sample. For N < 4, the uncertainties represent the two standard deviations (2 s.d.) of repeated analyses of the NIST 129c metal standard or the internal precision (2 s.e.), whichever is larger. The uncertainties for N ≥ 4 represent the Student-t 95% confidence intervals, i.e., ($t_{0.95,N-1}$ × s.d.)/√N.

Table 2

Platinum isotopic data for ungrouped iron meteorites and CB and ordinary chondrites.

| Sample | ID | N (Pt-IC) | $\varepsilon^{192}$Pt (6/5) (± 95% CI) | $\varepsilon^{194}$Pt (6/5) (± 95% CI) | $\varepsilon^{198}$Pt (6/5) (± 95% CI) | $\varepsilon^{192}$Pt (8/5) (± 95% CI) | $\varepsilon^{194}$Pt (8/5) (± 95% CI) | $\varepsilon^{196}$Pt (8/5) (± 95% CI) |
|---|---|---|---|---|---|---|---|---|
| **_Carbonaceous meteorites_** | | | | | | | | |
| **Iron meteorites** | | | | | | | | |
| Babb's Mill (Troost's Iron) | UI-01 | 7 | -0.57 ± 0.34 | -0.11 ± 0.04 | 0.17 ± 0.07 | -0.42 ± 0.34 | -0.05 ± 0.03 | -0.06 ± 0.02 |
| replicate | UI-01b | 6 | -1.17 ± 0.50 | -0.18 ± 0.06 | 0.17 ± 0.19 | -0.98 ± 0.64 | -0.11 ± 0.06 | -0.06 ± 0.06 |
| replicate | UI-01c | 6 | -0.52 ± 0.12 | -0.05 ± 0.06 | 0.12 ± 0.06 | -0.44 ± 0.15 | -0.01 ± 0.06 | -0.04 ± 0.02 |
| replicate | UI-01d | 6 | -1.01 ± 0.55 | -0.10 ± 0.08 | 0.21 ± 0.11 | -0.84 ± 0.58 | -0.04 ± 0.05 | -0.07 ± 0.04 |
| replicate | UI-01e | 6 | -0.70 ± 0.37 | -0.12 ± 0.05 | 0.19 ± 0.07 | -0.58 ± 0.30 | -0.07 ± 0.04 | -0.06 ± 0.02 |
| **mean** | | **5** | **-0.79 ± 0.35** | **-0.11 ± 0.06** | **0.17 ± 0.04** | **-0.65 ± 0.31** | **-0.06 ± 0.04** | **-0.06 ± 0.01** |
| Guffey | UI-20 | 3 | -0.80 ± 1.26 | -0.23 ± 0.17 | 0.33 ± 0.21 | -0.48 ± 1.23 | -0.12 ± 0.12 | -0.11 ± 0.07 |
| replicate | UI-20b | 3 | -0.68 ± 1.26 | -0.22 ± 0.17 | 0.27 ± 0.21 | -0.44 ± 1.23 | -0.12 ± 0.12 | -0.09 ± 0.07 |
| replicate | UI-20c | 3 | -0.40 ± 1.26 | -0.05 ± 0.17 | 0.15 ± 0.21 | -0.27 ± 1.23 | -0.02 ± 0.12 | -0.05 ± 0.07 |
| replicate | UI-20d | 3 | -0.23 ± 1.26 | -0.07 ± 0.17 | 0.13 ± 0.21 | -0.06 ± 1.23 | -0.03 ± 0.12 | -0.04 ± 0.07 |
| replicate | UI-20e | 4 | -0.50 ± 0.97 | -0.08 ± 0.08 | 0.14 ± 0.17 | -0.38 ± 0.97 | -0.05 ± 0.07 | -0.05 ± 0.06 |
| **mean** | | **5** | **-0.52 ± 0.28** | **-0.13 ± 0.11** | **0.20 ± 0.11** | **-0.33 ± 0.21** | **-0.07 ± 0.06** | **-0.07 ± 0.04** |
| Hammond | UI-21 | 1 | -1.37 ± 1.26 | -0.21 ± 0.17 | 0.42 ± 0.21 | -0.95 ± 1.23 | -0.06 ± 0.12 | -0.14 ± 0.07 |
| replicate | UI-21b | 2 | -0.66 ± 1.26 | -0.10 ± 0.17 | 0.18 ± 0.21 | -0.45 ± 1.23 | -0.02 ± 0.12 | -0.06 ± 0.07 |
| replicate | UI-21c | 1 | -0.67 ± 1.26 | -0.07 ± 0.17 | 0.11 ± 0.21 | -0.77 ± 1.23 | -0.06 ± 0.12 | -0.04 ± 0.07 |
| replicate | UI-21d | 1 | -1.39 ± 1.26 | -0.01 ± 0.17 | 0.09 ± 0.21 | -1.35 ± 1.23 | 0.01 ± 0.12 | -0.03 ± 0.07 |
| **mean[a]** | | **5** | **-0.95 ± 0.58** | **-0.10 ± 0.09** | **0.20 ± 0.20** | **-0.79 ± 0.62** | **-0.03 ± 0.05** | **-0.07 ± 0.07** |
| ILD 83500 | UI-22 | 5 | -0.63 ± 0.28 | -0.11 ± 0.08 | 0.18 ± 0.12 | -0.44 ± 0.21 | -0.05 ± 0.05 | -0.06 ± 0.04 |
| replicate | UI-22b | 6 | -0.75 ± 0.70 | 0.01 ± 0.11 | 0.01 ± 0.14 | -0.80 ± 0.70 | 0.03 ± 0.06 | 0.00 ± 0.04 |
| **mean** | | **2** | **-0.69 ± 0.17** | **-0.05 ± 0.17** | **0.10 ± 0.23** | **-0.62 ± 0.50** | **-0.01 ± 0.11** | **-0.03 ± 0.08** |
| **CB chondrites** | | | | | | | | |
| Gujba | AC02 | 7 | -0.25 ± 0.87 | -0.05 ± 0.06 | 0.12 ± 0.09 | -0.09 ± 0.87 | -0.01 ± 0.03 | -0.04 ± 0.03 |
| Isheyevo | AC03 | 5 | 0.94 ± 1.70 | -0.02 ± 0.06 | 0.07 ± 0.10 | 1.02 ± 1.60 | 0.01 ± 0.05 | -0.02 ± 0.03 |
| **Mean iron pre-exposure composition[b]** | | **4** | **-0.68 ± 0.13** | **-0.11 ± 0.04** | **0.17 ± 0.04** | **-0.47 ± 0.32** | **-0.05 ± 0.03** | **-0.06 ± 0.01** |
| **_Non-carbonaceous meteorites_** | | | | | | | | |
| **Ordinary chondrites** | | | | | | | | |
| Butsura | AC04 | 5 | 0.69 ± 0.36 | 0.11 ± 0.09 | -0.08 ± 0.13 | 0.63 ± 0.25 | 0.10 ± 0.05 | 0.03 ± 0.04 |
| NWA 6630 | AC06 | 3 | 1.00 ± 1.26 | 0.10 ± 0.17 | -0.16 ± 0.21 | 0.80 ± 1.23 | 0.05 ± 0.12 | 0.05 ± 0.07 |
| NWA 6629 | AC07 | 2 | 0.58 ± 1.26 | 0.11 ± 0.17 | -0.07 ± 0.21 | 0.51 ± 1.23 | 0.07 ± 0.12 | 0.02 ± 0.07 |

Platinum isotopic ratios are normalized to $^{196}$Pt/$^{195}$Pt = 0.7464 (6/5) and $^{198}$Pt/$^{195}$Pt = 0.2145 (8/5) (Kruijer et al., 2013). N is the number of measurements per sample. For N < 4, the uncertainties represent the two standard deviations (2 s.d.) of repeated analyses of the NIST 129c metal standard or the internal precision (2 s.e.), whichever is larger. The uncertainties for N ≥ 4 represent the Student-t 95% confidence intervals, _i.e._, (t$_{0.95,N-1}$ × s.d.)/√N.

[a] Average (±95% CI) of five individual analysis from four digestions.

[b] Weighted average (±95% CI) of Babb's Mill, Guffey, Hammond, and ILD 83500.

Table 3

Revised pre-exposure ε$^{182}$W values and corresponding Hf-W model ages of core formation for the major magmatic iron meteorite groups.

| Iron meteorite group | ε$^{182}$W (6/4) (± 95% CI) | Differentiation age [Ma] (± 95% CI) |
|---|---|---|
| **NC iron meteorites** | | |
| IC | -3.37 ± 0.04 | 1.0 ± 0.7 |
| IIAB | -3.32 ± 0.03 | 1.4 ± 0.7 |
| IIIAB | -3.27 ± 0.03 | 1.9 ± 0.7 |
| IIIE | -3.20 ± 0.06 | 2.6 ± 0.9 |
| IVA | -3.24 ± 0.05 | 2.2 ± 0.8 |
| **CC iron meteorites** | | |
| IIC | -3.12 ± 0.12 | 3.4 ± 1.4 |
| IID | -3.15 ± 0.04 | 3.1 ± 0.7 |
| IIF | -3.13 ± 0.05 | 3.3 ± 0.8 |
| IIIF | -3.16 ± 0.10 | 3.0 ± 1.2 |
| IVB | -3.10 ± 0.05 | 3.6 ± 0.8 |

W isotope data from Kruijer et al. (2014, 2017), Hilton et al. (2019), and Tornabene et al. (2020). Hf-W model ages calculated using equation 3 and given in Ma after CAI formation.

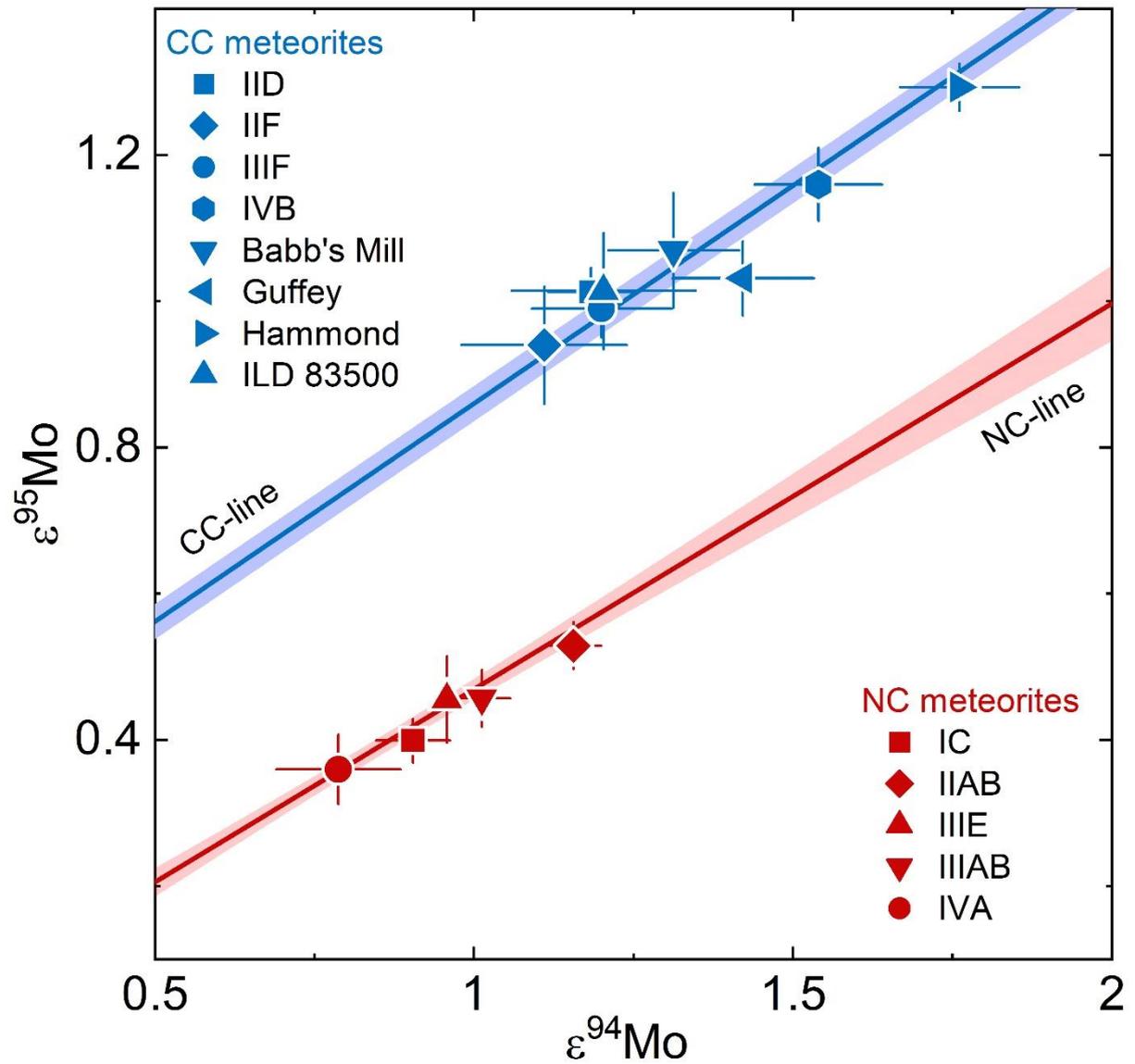

**Figure 1.** Diagram of $\varepsilon^{95}$Mo vs. $\varepsilon^{94}$Mo for iron meteorites. The ungrouped irons from this study are shown together with data for the major groups of iron meteorites as summarized in Spitzer et al. (2020). NC- and CC-lines as defined by Spitzer et al. (2020) and Budde et al. (2019), respectively.

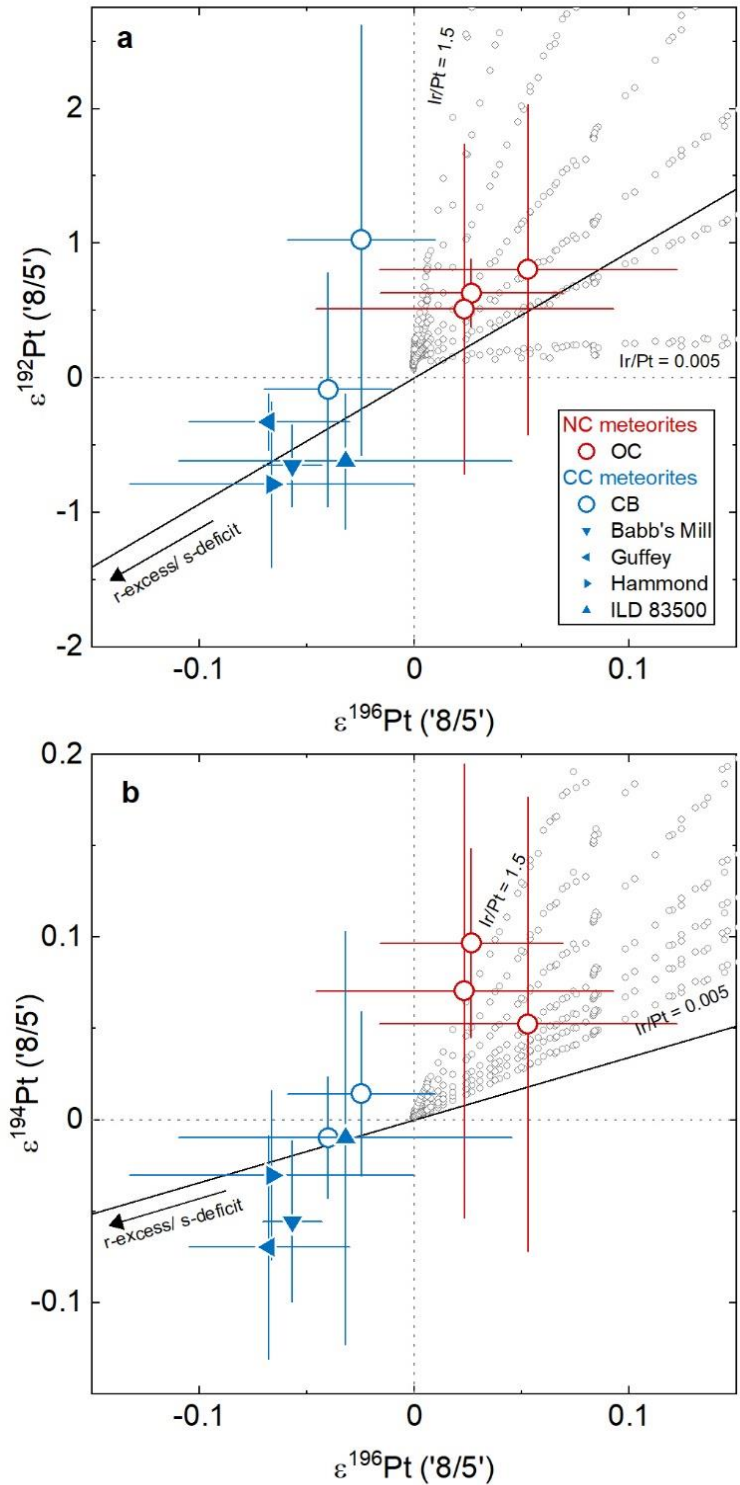

**Figure 2.** (a) Diagram of $\varepsilon^{192}$Pt vs. $\varepsilon^{196}$Pt and (b) $\varepsilon^{194}$Pt vs. $\varepsilon^{196}$Pt for the samples of this study. Shown as small grey circles are modeled effects of neutron capture in iron meteorites for different Ir/Pt (Kruijer et al., 2013; Leya and Masarik, 2013). Black solid lines represent *s-/r-*process mixing lines calculated using the stellar model from Arlandini et al. (1999).

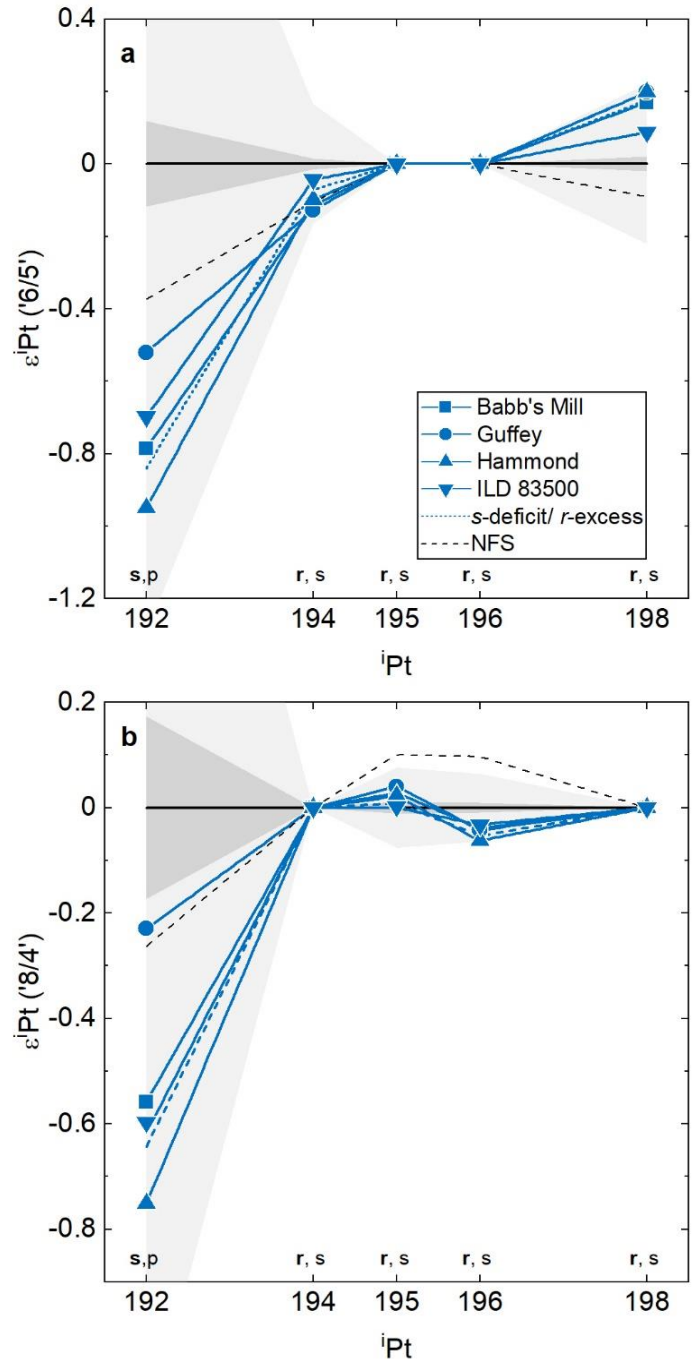

**Figure 3.** Pt isotope anomalies of the four ungrouped iron meteorites of this study normalized to $^{198}$Pt/$^{195}$Pt (a) and $^{198}$Pt/$^{194}$Pt (b). The isotope pattern of these samples is consistent with a deficit in *s*-process or excess in *r*-process Pt isotopes (blue dashed line), but not with fractionation due to the nuclear field shift (NFS) effect. Nucleosynthetic pathways for each isotope are indicated on the *x*-axis. The light and dark shaded areas represent the 2.s.d. and 95% CI determined by repeated measurements of the NIST 129c steel.

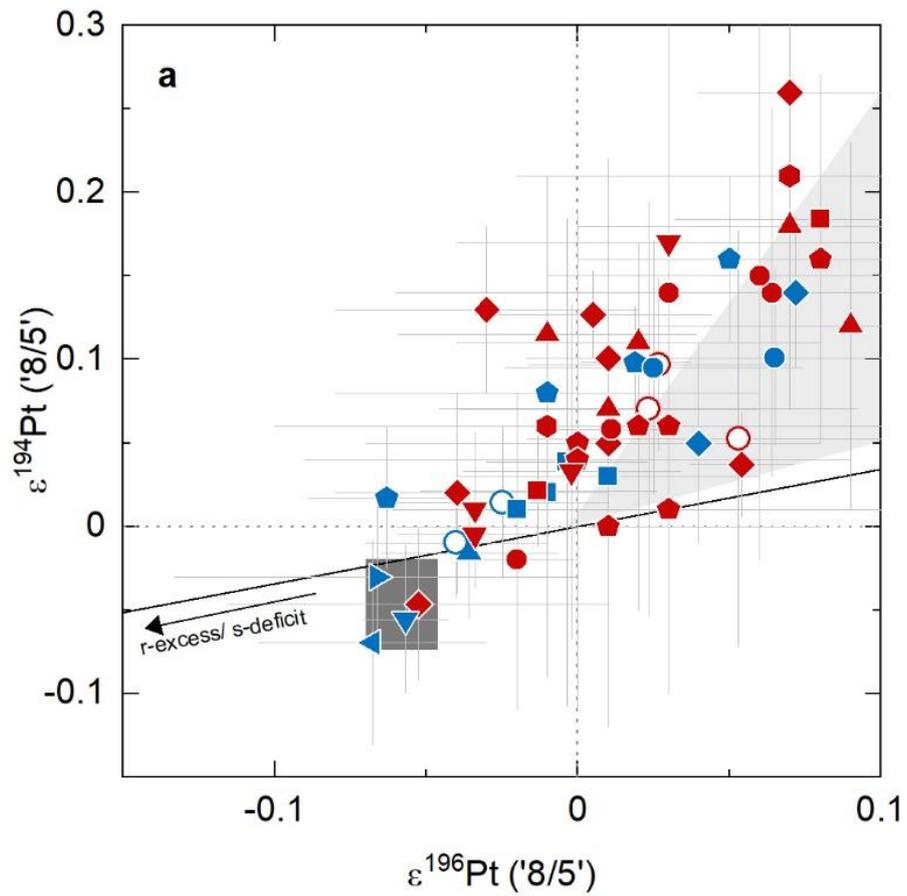

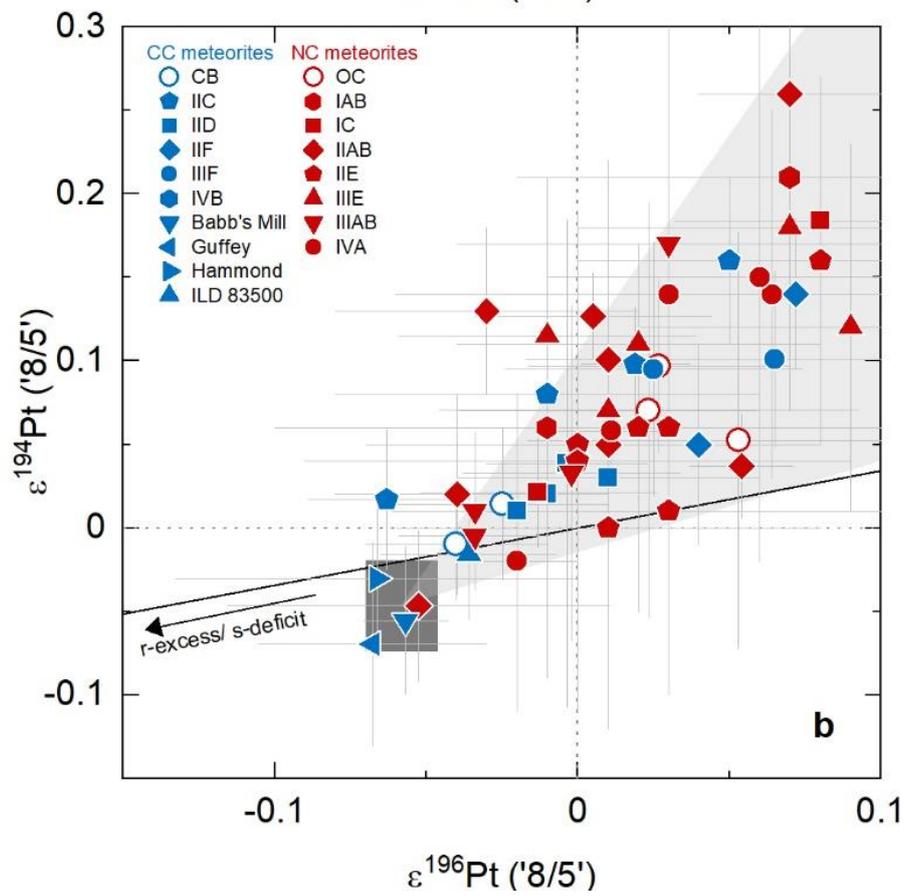

**Figure 4.** Diagram of $\varepsilon^{194}$Pt vs. $\varepsilon^{196}$Pt for the samples of this study together with additional iron meteorite literature data (Table S6). Dark grey rectangle represents the pre-exposure Pt isotope composition of iron meteorites determined in this study (Table 2). Light grey area indicates modelled CRE effects. (a) For a pre-exposure Pt isotope composition of $\varepsilon^{194}$Pt = $\varepsilon^{196}$Pt = 0, several iron meteorites plot outside the modelled field of CRE effects (light grey area). (b) If instead the Pt isotope composition of the four ungrouped irons from this study is used as the pre-exposure composition, all samples plot inside the field of calculated CRE effects for Ir/Pt ratios between 0.005 and 1.5, which correspond to the total range reported for these samples (Kruijer et al., 2013; Leya and Masarik, 2013). Black solid line represents *s*-/*r*-process mixing line as in Fig. 2.

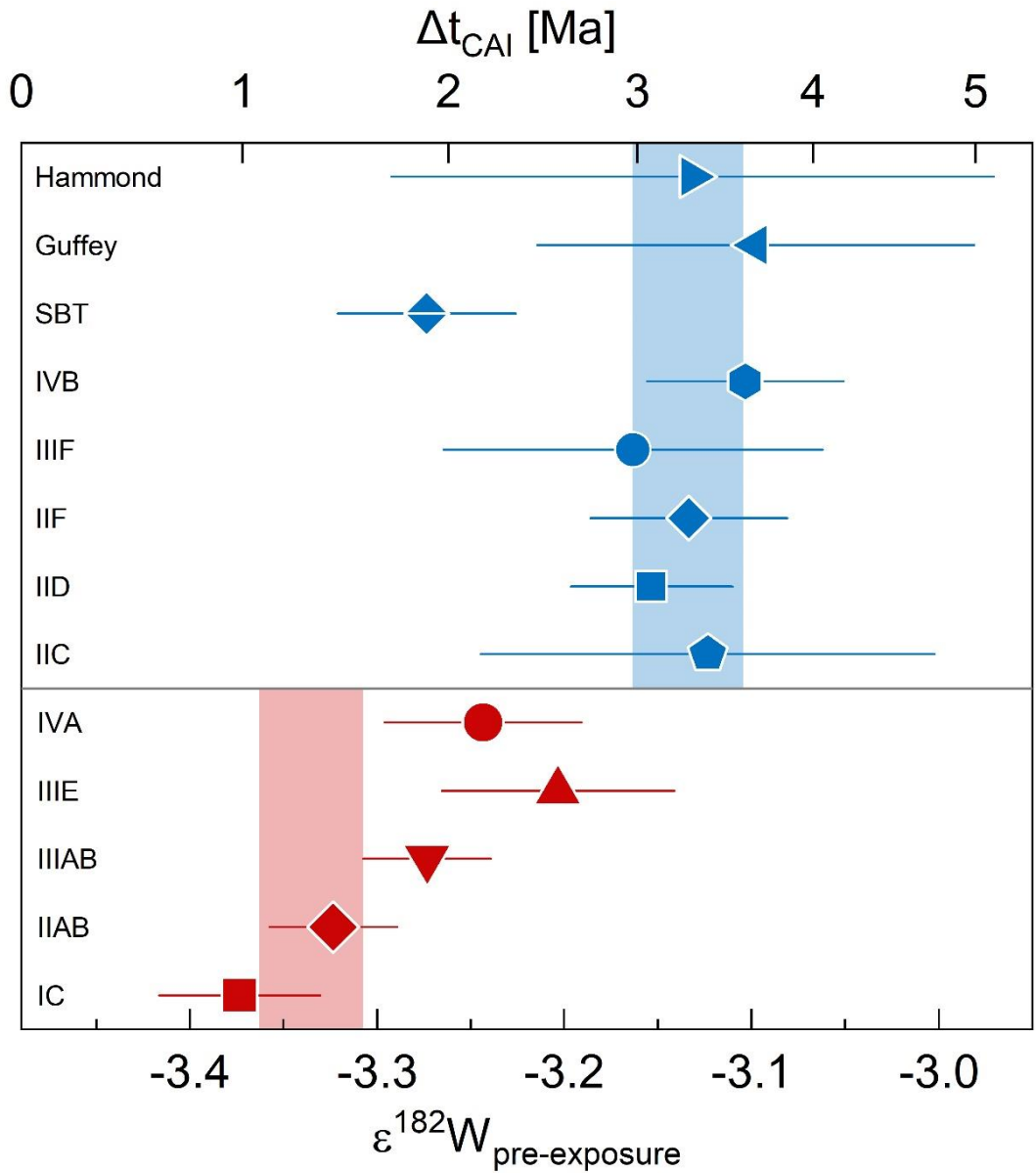

**Figure 5**. Revised Hf-W model ages for iron meteorites after correction for a nucleosynthetic Pt isotope variations. Red and blue shaded areas represent the mean ages of the volatile-rich NC (IC, IIAB) and CC iron groups (except SBT), respectively. Hammond and Guffey together with all CC groups have indistinguishable core formation ages of ~3.3 Ma after CAIs, whereas the SBT (includes ILD 83500 and Babb's Mill) displays a slightly older age, which overlaps with those of NC irons.

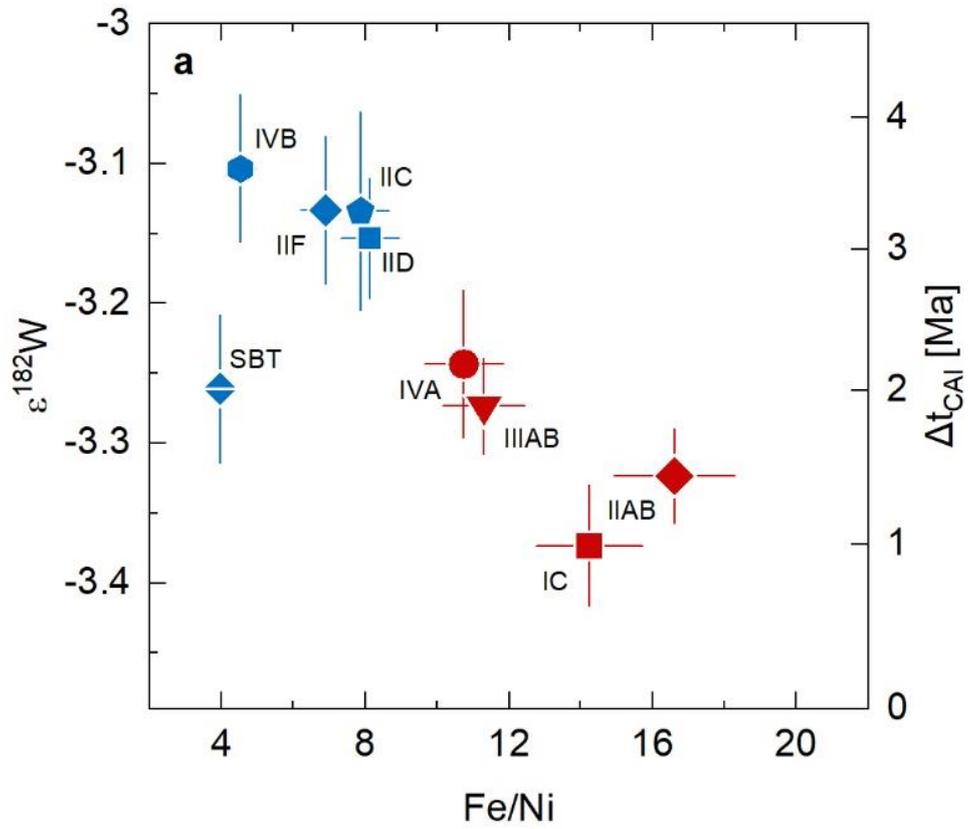

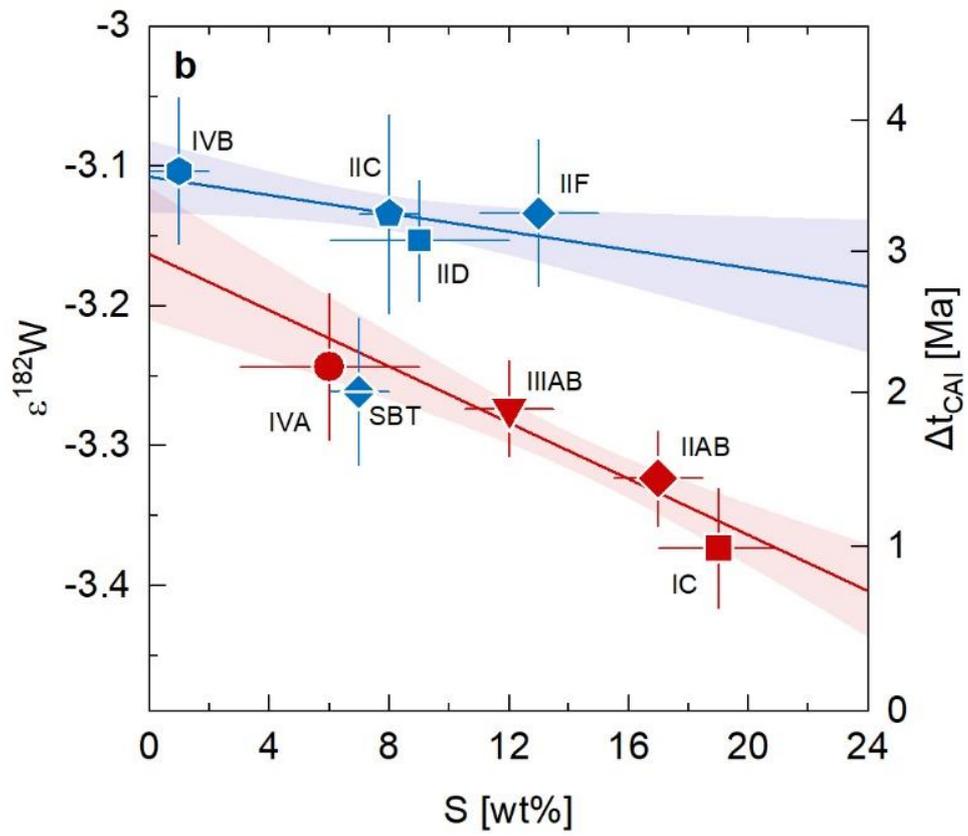

**Figure 6.** Diagrams of $\varepsilon^{182}W$ versus (a) Fe/Ni ratio and (b) S contents for bulk cores of the major magmatic iron meteorite groups and the South Byron Trio. Fe/Ni ratios and S contents as summarized in Table S7. The blue and red lines are York-fits with 1 $\sigma$ error envelopes fitted through the NC and CC (except SBT) iron groups. CC irons have systematically lower Fe/Ni ratios compared to NC irons, indicating more oxidizing conditions of core formation in CC bodies. In spite of their variable S contents, all CC irons (except SBT) have indistinguishable $\varepsilon^{182}W$ values, suggesting that the depletion of S (and probably other volatile elements) occurred after core formation, which in turn implies that the CC iron meteorite parent bodies were volatile-richer than they are today.

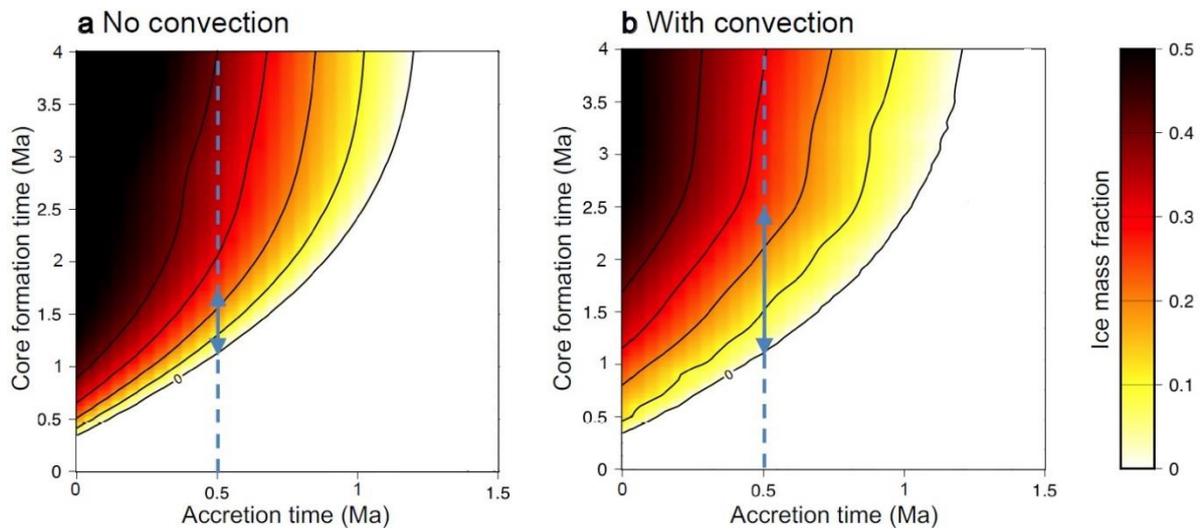

**Figure 7.** Results of thermal models for a 40 km radius asteroid with 1.2 wt.-% Al, showing core formation time as a function of accretion time and water ice mass fraction (see supplement). (a) Melt-water is static. (b) Melt-water is assumed to convect ($K_{ref} = 5 \times 10^{-11} m^2$), increasing the rate of heat transfer. Dashed lines and arrows indicate how changing water ice mass fraction from 0 to 25% changes the core formation time for an accretion time of 0.5 Ma.